\documentclass[conference]{IEEEtran}
\usepackage{cite}
\ifCLASSINFOpdf
  \usepackage[pdftex]{graphicx}
\else
\fi
\usepackage[cmex10]{amsmath}
\usepackage{amssymb}
\usepackage{amsthm}
\usepackage{url}
\usepackage{multirow}	% VL fusionner des cellules verticalement dans les tables
\usepackage[bookmarks=true, bookmarksnumbered=true, bookmarksopen=true,	% VL générer un PDF avec des hyperliens
		unicode=true, colorlinks=true,
		linkcolor=blue,citecolor=blue,filecolor=blue,urlcolor=blue
		]{hyperref}

{\bfseries}{\itshape}

\begin{document}
\title{Identifying the Community Roles of Social Capitalists in the Twitter Network}

\author{\IEEEauthorblockN{Vincent Labatut}
\IEEEauthorblockA{Galatasaray University, Computer Science Department \\ \c C{\i}ra\u{g}an cad. n°36, Ortak\"oy 34357, \\ \.{I}stanbul, Turquie}
\and
\IEEEauthorblockN{Nicolas Dugu\'e, Anthony Perez}
\IEEEauthorblockA{Univ. Orl\'eans, INSA Centre Val de Loire \\ LIFO EA 4022 45067 Orl\'eans, France}}


%Telephone: (800) 555--1212, Fax: (888) 555--1212}


\maketitle

\begin{abstract}
%\boldmath
In the context of Twitter, social capitalists are specific users trying to increase their number of followers and interactions by any means. These users are not healthy for the Twitter network since they flaw notions of influence and visibility. Indeed, it has recently been observed that they are real and active users that can help malicious users such as spammers gaining influence. Studying their behavior and understanding their position in Twitter is thus of important interest. A recent work provided an efficient way to detect social capitalists using two simple topological measures. Based on this detection method, we study how social capitalists are distributed over Twitter's friend-to-follower network. We are especially interested in analyzing how they are organized, and how their links spread across the network. Answering these questions allows to know whether the social capitalism methods increase the actual visibility on the service. To that aim, we study the position of social capitalists on Twitter w.r.t. the community structure of the network. % DONNER L'INTERET.
%The notion of community structure is particularly useful because it provides an intermediate level, compared to the more classic global (whole network) and local (node neighborhood) approaches. 
We base our work on the concept of community role of a node, which describes its position in a network depending on its connectivity at the community level. The topological measures originally defined to characterize these roles 
%were designed for undirected networks, they 
consider only some aspects of community-related connectivity and rely on a set of empirically fixed thresholds. We first show the limitations of such measures  and then extend and generalize them by considering new aspects of the community-related connectivity. Moreover, we use an unsupervised approach to distinguish the roles, in order to provide more flexibility relatively to the studied system. We then apply our method to the case of social capitalists and show that they are highly visible on Twitter, due to the specific roles they occupy.
\end{abstract}


% \fi
\IEEEpeerreviewmaketitle
\section{Introduction}

%A complex network is a graph representing a real-world complex system. Such a system possesses emerging properties, which are caused by the way its constituting elements interact. This, in turns, gives non-trivial topological properties to the graph modeling the system. During the last fifteen years, various such properties have been identified, such as being small-world \cite{Watts1998}, being scale-free \cite{Barabasi1999a} or having a community structure \cite{Girvan2002a}. In this article, we focus on the latter, known to appear in many real-world systems \cite{Fortunato2010}.

\noindent \textbf{Context.} The last decade has been marked by an increase in both the number of online social networking services and the number of users of such services. This observation is particularly relevant when considering Twitter, which had $200$ millions accounts in April $2011$~\cite{Bosker2011} and reached $500$ millions accounts in October $2012$~\cite{Holt2013}. 
Twitter is mostly used to share, seek and debate about information, or to let the world know about daily events~\cite{Java2007}. The amount of information shared on Twitter is considerable: there are about $1$ billion tweets posted every two and a half days~\cite{Rodgers13}. While focusing on microblogging, Twitter can be considered as a social networking service, since it includes social features. Indeed, to see the messages of other users, a Twitter user has to \emph{follow} them (i.e. make a subscription). Furthermore, a user can \emph{retweet}~\cite{Suh2010} other users' tweets, for instance when he finds them interesting and wants to share them with their followers. Besides, users can \emph{mention} other users to draw their attention by adding \texttt{@UserName} in their message. %With all these features, Twitter allows interaction between its users. 
Some Twitter users are trying to use these particular properties to spread efficiently some information~\cite{GVK+12}.
One of the simplest way to reach this objective is to gain as many followers as possible, since this gives a higher visibility to the user's tweets when using the network search engines~\cite{GVK+12}.

\noindent \textbf{Social capitalists.}  These specific users are called \emph{social capitalists}. They have been recently observed and studied by Ghosh \textit{et al}.~\cite{GVK+12} in a study related to \emph{link-farming} in Twitter. They noticed in particular that users responding the most to the solicitation of spammers are in fact \emph{real}, \emph{active} users. To increase their number of followers, social capitalists use several techniques~\cite{DP14,GVK+12}, the most common one being to follow a lot of users \emph{regardless of their content}, just hoping to be followed back.
%use two distinct techniques called \textit{I follow you, follow me} (IFYFM) and \textit{Follow me, I follow you} (FMIFY). The former consists in following many users in the hope of being followed back in return, whereas in the latter the social capitalist promises other users he will follow them back if they follow him first. 
Because of this lack of interest in the content produced by the users they follow, social capitalists are not healthy for a service such as Twitter. Indeed, this behavior helps spammers gaining influence~\cite{GVK+12}, and more generally makes the task of finding relevant information harder for regular users. Studying their behavior and understanding their position in Twitter is therefore a very important task to improve the service, since it can allow designing better search engines or functioning rules. In a recent work, Dugu\'e and Perez~\cite{DP14} have shown that social capitalists can be efficiently detected and classified using two purely topological measures, called \emph{overlap}~\cite{Sim43} and \emph{ratio} indices. They provide useful information regarding the interaction between the set of \emph{friends} and the set of \emph{followers}\footnote{For a given user, \emph{friends} denote the set of users he follows, and \emph{followers} the set of users that follow him, as per the official Twitter terminology.} of a user, which are supposed to have a large intersection whenever a user applies social capitalism techniques. In this work, we rely on this detection method to characterize the behavior of social capitalists. To better understand how they are organized, how really visible they are and how their links spread across the network, we study the positions that social capitalists occupy in Twitter w.r.t. the community structure of the network. % DONNER L'INTERET.
%The notion of community structure is particularly useful because it provides an intermediate level, compared to the more classic global (whole network) and local (node neighborhood) approaches. 

\noindent \textbf{Community roles.} 
In its simplest form, the community structure of a complex network can be defined as a partition of its node set, each part corresponding to a community. Community detection methods generally try to perform this partition in order to obtain densely connected groups of nodes, relatively to the rest of the network \cite{Newman2004a}. Hundreds of such algorithms have been defined in the last ten years, see \cite{Fortunato2010} for a very detailed review of the domain. The notion of community structure is particularly interesting because it allows studying the network at an intermediate level, compared to the more classic global (whole network) and local (node neighborhood) approaches.

The concept of community role is a good illustration of this characteristic. It consists in describing a node depending on the position it holds in its own community\footnote{Note that the notion of role also appears in works related to block modeling, but it is not defined in terms of position in a community~\cite{Guimera2005}.}.  Community roles were initially introduced by Guimer\`a and Amaral~\cite{Guimera2005} to study metabolic networks. After having applied a standard community detection method, they characterize each node according to two \textit{ad hoc} measures, each one describing a specific aspect of the community-related connectivity. 
%The first expresses the intensity of its connections to the rest of its community, whereas the second quantifies how uniformly it is connected to all communities. 
The node role is then selected amongst $7$ predefined ones by comparing the two values to some empirically fixed thresholds. Guimer\`a and Amaral~\cite{Guimera2005} showed certain systems possess a role invariance property: when several instances of the system are considered, nodes are different but roles are similarly distributed. Scripps \textit{et al}.~\cite{Scripps2007}, apparently unaware of this previous work, later adopted a similar approach, but this time for influence maximization and link-based classification purposes. They also use two measures: first the degree, to assess the intensity of the general node connectivity, and second an \textit{ad hoc} measure, to reflect the number of communities to which it is connected. They then use arbitrary thresholds to define $4$ distinct roles. 
%Finally, the embeddedness measure defined by \cite{Lancichinetti2010} can be considered as a related work, even if their goal was not to study community roles. Indeed, it captures how much a node is integrated to its community, by considering the proportion of its neighbors located in the same community.

\noindent \textbf{Our contribution.} In this paper, we study the community roles of social capitalists within a freely-available Twitter network provided by Cha \textit{et al}.~\cite{CHBG10}. We focus on the concept of community role as described by~Guimer\`a and Amaral~\cite{Guimera2005}, because it relies more heavily on the community structure. In a first place, we highlight two important limitations of this community role approach. We show that the existing measures do not take into account all aspects of the community-related external connectivity of a node. Moreover, we object the assumption of universality of the thresholds applied to the measures in order to distinguish the different node roles. The dataset we use constitutes a counter-example showing the original thresholds are not relevant for all systems. 
%Additionally, this approach was designed to handle only undirected links, so this type of information is ignored even when it is available, as it is in our case.
We then explain how to tackle these limitations. We first introduce three new measures to characterize the external connectivity of a node in a more complete and detailed way. We then describe an unsupervised approach aiming at identifying the node roles without using fixed thresholds. Finally, we apply our method on the Twitter network to determine the position of social capitalists, and show they occupy specific roles in the network. 
In particular, most of them are well connected to their community, and overall a large part of them spread their links outside their community very efficiently. This gives meaningful insights regarding the actual visibility of these users. They thus seem to occupy roles leading to a high visibility in Twitter. 

\noindent \textbf{Outline.} We first present the concept of social capitalists in Twitter in more details (Section~\ref{sec:capsoc}). Next, we describe the method proposed by~Guimer\`a and Amaral~\cite{Guimera2005} to identify the community roles of nodes (Section~\ref{subsec:original}) and provide some elements towards its limitation (Section~\ref{subsec:participationlimits}). We then describe the solutions we propose to tackle these limitations (Section~\ref{subsec:generalizedmeasures}) and finally apply our method to study the roles of social capitalists in Twitter (Section~\ref{sec:results}). 

%In this work, we explain how to tackle these limitations. First, we adapt the analysis method to directed networks. Second, we define additional measures allowing to consider each aspect of the community-related connectivity of a node: community diversity, heterogeneity of link distribution, and intensity of the connection. Third, we propose an automatic unsupervised approach to determine roles based on the processed measures. Finally, we illustrate the applicability of our method by analyzing a network of followee-to-follower relationships on Twitter~\cite{CHBG10}. We show how our proposed modifications allow discovering the fact some particular users, called social capitalists~\cite{DP14,GVK+12}, occupy very specific roles in this system.
%
%The rest of the article is organized as follows. In section~\ref{sec:original}, we describe and discuss the approach proposed in Guimer\`a and Amaral~\cite{Guimera2005}. In section \ref{sec:proposed}, we detail the modifications we propose. In section \ref{sec:results}, we present our real-world data and the concept of social capitalist, before describing the results obtained with our method. Section \ref{sec:conclusion} summarizes our contribution and opens some perspectives.

\section{Social Capitalists in Twitter}
\label{sec:capsoc}
Social capitalists have first been highlighted by Ghosh \textit{et al}.~\cite{GVK+12} during a study focused on \emph{link-farming} and \emph{spammers} in Twitter. These specific Twitter users try to increase their number of followers by any means. To achieve this goal, they exploit two relatively straightforward principles based on the reciprocation of the \emph{follow} link:

\begin{itemize}
	\item[-] \textbf{FMIFY} (Follow Me and I Follow You): the user ensures his potential followers that he will follow them back if they follow him first; 
	\item[-] \textbf{IFYFM} (I Follow You, Follow Me): on the contrary, the user systematically follows other users, hoping to be followed back.
\end{itemize}

In their work, Ghosh \textit{et al.}~\cite{GVK+12} noticed that users responding the most to the solicitations of spammers are real (\emph{i.e.} neither bots nor fake accounts), \emph{active} and even sometimes \emph{popular} users, that they called \emph{social capitalists}. Using this observation, they constituted a list of $100,000$ social capitalists -namely the most responsive ones to the solicitations of spammers. 
Social capitalists are not healthy for a social networking service, since their methods to gain visibility and influence are not based on the production of relevant content and on getting a higher credibility. From this point of view, their high number of followers can be considered as undeserved, and biases all services based on the assumption that visible users produce or fetch interesting content (e.g. search or recommendation engines).

Using two purely topological measures (and therefore without considering any content), Dugu\'e and Perez~\cite{DP14} designed a method to detect and classify efficiently these users. These measures are based on neighborhood comparisons, namely between the sets of followers $N^-(u)$ (incoming neighbors) and friends $N^+(u)$ (outgoing neighbors) of a user of interest $u$. The first is called the \textit{overlap index}~\cite{Sim43}, and is used to detect social capitalists: 
\begin{eqnarray}	
	O(u) = \frac{|N^-(u) \cap N^+(u)|}{\min{\{|N^-(u)|,|N^+(u)|\}}}
\label{eq:overlap}		
\end{eqnarray}
\noindent Its value ranges from $0$ (regular user) to $1$ (social capitalist). The second is the \textit{ratio} $r$, and is used to distinguish between social capitalists using the \textbf{FMIFY} ($r \leq 1$) and \textbf{IFYFM} ($r>1$) techniques: 
\begin{eqnarray}	
	r(u) = \frac{|N^+(u)|}{|N^-(u)|}
\label{eq:ratio}		
\end{eqnarray}

Dugu\'e and Perez~\cite{DP14} also use a third criterion, the number of followers, which corresponds to the incoming degree of the considered node, noted $d^{in}(u)$. 
They define \emph{low in-degree social capitalists} as social capitalists having less between $500$ $10,000$ followers, and \emph{high in-degree social capitalists} as the remaining set of social capitalists. The latter users are considered as successful social capitalists, while the former ones are more popular. It is interesting to notice that in the network we consider, most users with more than $10,000$ followers are social capitalists ($70\%$). Moreover, users with such a number of followers constitute less than $0.1\%$ of the network, which justifies their popularity. 
%Dugu\'e and Perez consider the social capitalists have a \textit{low} in-degree when it is between $500$ and $10,000$, and a \emph{high} one when it is greater than $10,000$. The latter are efficiently gaining followers, whereas the former are still less popular.

In the experimental part of this article, we decide to use this method to identify the social capitalists in the studied data, instead of the list manually curated by~\cite{GVK+12}. The reason for this is that the latter seems less exhaustive since it excludes users who do not follow spammers, and does not contain spammers nor bots. Furthermore, some of them have only a few followers, or only a few reciprocate followers-friends links. Finally, the method proposed by Dugu\'e and Perez~\cite{DP14} achieved a greater than $80\%$ accuracy when comparing the social capitalists it detected with those from the list. 

\section{Identifying Community Roles} 
\label{sec:communityroles}
%We now present in more details the concept of community role in a complex network. We first introduce the original measures of Guimer\`a and Amaral~\cite{Guimera2005}, then highlight their limitations, and finally propose some solutions to these problems.
In order to characterize the roles of nodes in communities, Guimer\`a and Amaral~\cite{Guimera2005} defined two complementary measures which allow them to place each node on a 2D role space. Then, they proposed various thresholds to discretize this space, each resulting subspace corresponding to a specific role. We first present this method, then highlight its limitations, and finally propose some solutions to these problems. 

\subsection{Original approach}
\label{subsec:original}		
%Our work relies on the method proposed by~Guimer\`a and Amaral~\cite{Guimera2005}, not only because it is much more widespread than that by~\cite{Scripps2007}, but also because it relies more heavily on the community structure. 

%In this section, we first describe the measures, then the method used to identify the roles.

\noindent \textbf{Measures.} The two measures are related to the \textit{internal} and \textit{external connectivity} of the node with respect to its community. In other words, they respectively deal with how a node is connected with other nodes inside and outside of its own community. The first measure, called \emph{within-module degree}, is based on the notion of \emph{$z$-score}. Since the $z$-score will be used again afterwards, we define it in a generic manner. Let $f(u)$ be any function defined on the nodes, that is $f$ associates a numerical value to any node $u$ of the considered graph. The $z$-score $Z_f(u)$ w.r.t. the community of $u$ is defined by:
\begin{eqnarray}	
	Z_f(u) & = & \frac{f(u) - \mu_i(f)}{\sigma_i(f)}, u \in C_i	
\label{eq:zscore}		
\end{eqnarray}
\noindent where $C_i$ stands for a community, and $\mu_i(f)$ and $\sigma_i(f)$ respectively denote the mean and the standard deviation of $f$ over the nodes belonging to community $C_i$. 

Now, let $d_{int}(u)$ be the \textit{internal degree} of a node $u$, i.e. the number of links $u$ has with nodes belonging to its own community. Then, the \textit{within-module degree} of a node $u$, denoted $z(u)$ by Guimer\`a and Amaral~\cite{Guimera2005}, corresponds to the $z$-score of its internal degree. Note that $z$ evaluates the connectivity of a node towards its community with respect to that of the other nodes of the same community. 

The second measure, called \emph{participation coefficient}, is defined as follows:
\begin{eqnarray}
	P(u) & = & 1 - \sum_i \Big ( \frac{d_i(u)}{d(u)} \Big )^2
\label{eq:participation}
\end{eqnarray}
\noindent where $d(u)$ denotes the \emph{degree} of the node (\emph{i.e.} the number of links it has towards other nodes), and $d_i(u)$ the \textit{community degree} of $u$ (i.e. the number of links it has towards nodes of community $C_i$). Note that when $C_i$ corresponds to the community of $u$, then $d_i(u) = d_{int}(u)$. Roughly speaking, the participation coefficient evaluates the connectivity of a node to the communities. If it is close to $0$, then the node is connected to one community only (likely its own). On the contrary, if it is close to $1$, then the node is uniformly linked to a large number of communities.

\noindent \textbf{Community Roles.} Both measures are used to characterize the \emph{role} of a node within its community. Guimer\`a and Amaral~\cite{Guimera2005} defined $7$ different roles by discretizing the 2D space formed by $z$ and $P$. They first used a threshold on the within-module degree, which allowed them to distinguish \emph{hubs} (that is, nodes with $z \geqslant 2.5$) from other nodes (called non-hubs). Such hubs are considered as highly linked to their community, when compared to other nodes of the same community. Those two categories are subdivided thanks to several thresholds defined on the participation coefficient (by order of increasing $P$), as shown in Table~\ref{tab:roleDesc}.

\begin{table}[h]
	\centering
	\begin{tabular}{|l|l|l|l|l|}
		\hline
		\multicolumn{4}{|c|}{\textbf{Community role}} & \multicolumn{1}{|c|}{\textbf{External}} \\
		\cline{1-4}
		\multicolumn{2}{|c|}{\textbf{Within-Module Degree}} & \multicolumn{2}{|c|}{\textbf{Participation Coefficient}} & \multicolumn{1}{|c|}{\textbf{Connectivity}} \\
		\hline
		\multirow{3}{*}{Hub} & \multirow{3}{*}{$z \geq 2.5$} & Provincial & $P \leq 0.30$ & Low \\
		& & Connector & $P \in ]0.30;0.75]$ & Strong \\ %
		& & Kinless & $P > 0.75$ &  Very strong \\
		\hline
		\multirow{3}{*}{Non-Hub} & \multirow{3}{*}{$z<2.5$} & Ultra-peripheral & $P \leq 0.05$ & Very low  \\
		& & Peripheral & $P \in ]0.05;0.62]$ & Low  \\
		& & Connector & $P \in ]0.62;0.80]$ &   Strong \\
		& & Kinless & $P > 0.80$ &  Very strong \\
		\hline
	\end{tabular}\\[0.4cm] 
	\caption{Classification of roles according to their community-related connectivity.}
	\label{tab:roleDesc}
\end{table}

\noindent \textbf{Directed Variants.}
\label{sec:directedmeasures}
Many networks representing real-world systems, such as the Twitter network we study here, are directed. Of course, it is possible to analyze them through the undirected method, but this would result in a loss of information. %The question is to know whether this information is valuable to our objective. To check this matter, we extended the original measures in order to deal with this additional information. A comparison between the results obtained with the undirected and directed version will allow us to assess the importance of link direction in the detection of roles. 

Yet, extending these measures is quite straightforward: the standard way of proceeding consists in distinguishing incoming and outgoing links. In our case, this results in using $4$ measures instead of $2$: in- and out- versions of both the within-module degree and participation coefficient. Let us note $d^{in}$  the in-degree of a node, i.e. the number of incoming links connected to the node. Then one can consider the \textit{internal in-degree} of a node, noted $d^{in}_{int}$, corresponding to the number of incoming links the node has inside its community. By processing the $z$-score of this value, one can derive the \textit{within-module in-degree}, noted $z^{in}$. Let us note $d^{in}_i$  the \textit{community in-degree}, i.e. the number of incoming links a node has from nodes in community $C_i$. We can now define the \textit{incoming participation coefficient}, noted $P^{in}$, by substituting $d^{in}$ to $d$ and $d^{in}_i$ to $d_i$ in Equation (\ref{eq:participation}). We similarly define $z^{out}$ and $P^{out}$, using the outgoing counterparts $d^{out}$, $d^{out}_{int}$ and $d^{out}_i$. In the rest of the article, we call this set of measures the \textit{directed variants}, by opposition to the \textit{original measures} of Guimer\`a and Amaral~\cite{Guimera2005}.

%It is worth noticing it would not be consistent to process such directed measures on communities identified while ignoring link directions. So, it is necessary to apply a community detection method able to take this information into account. Another important point is the definition of roles. There is no reason the thresholds defined by Guimer\`a and Amaral~\cite{Guimera2005} are still valid for their directed variants. Fortunately, the unsupervised method we propose in section \ref{sec:unsupervised} tackles this problem.

%TODO AP est-ce qu'on dit ici que c'est non-orienté... ?
%We believe that several aspects of this approach can be discussed. As mentioned previously, the threshold are defined empirically, without any prior knowledge of the network, but are claimed to be universal. Next, we believe that the notion of participation coefficient is not accurate enough. Indeed, TODO. We use the notion of social capitalists to enlight these observations.  

\subsection{Limitations of this approach} %%Participation limit
% TODO AP a enlever si on manque de place, on peut s'en passer je pense
As mentioned before, we identify two limitations in the approach of Guimer\`a and Amaral~\cite{Guimera2005}. The first concerns the way the participation coefficient represents the nodes external connectivity, whereas the second is related to the thresholds used for the within-module degree.

\begin{figure}[h]
	\begin{minipage}{0.32 \linewidth}
		\centerline{\includegraphics[scale=0.10]{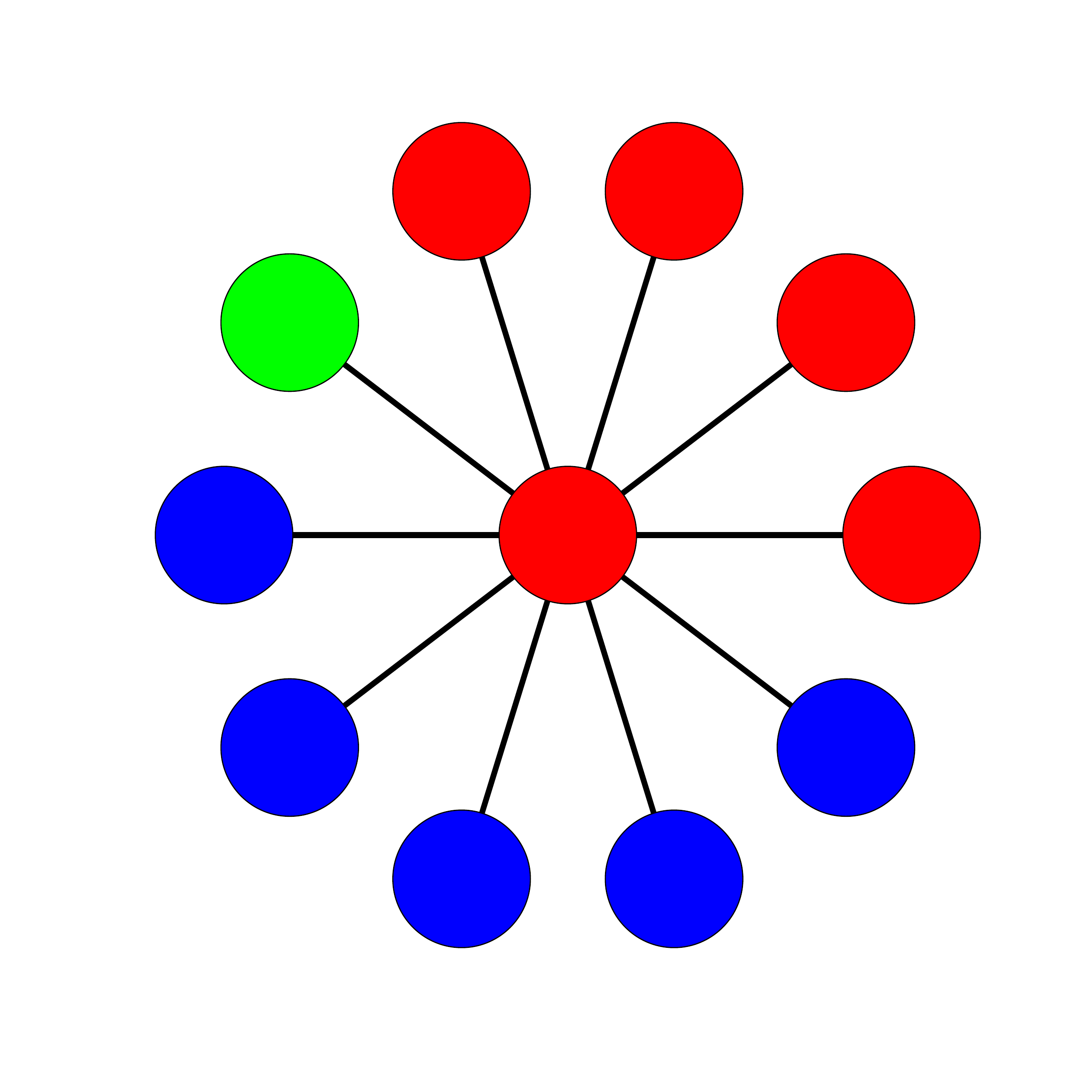}}
	\end{minipage}
	\begin{minipage}{0.32 \linewidth}
		\centerline{\includegraphics[scale=0.10]{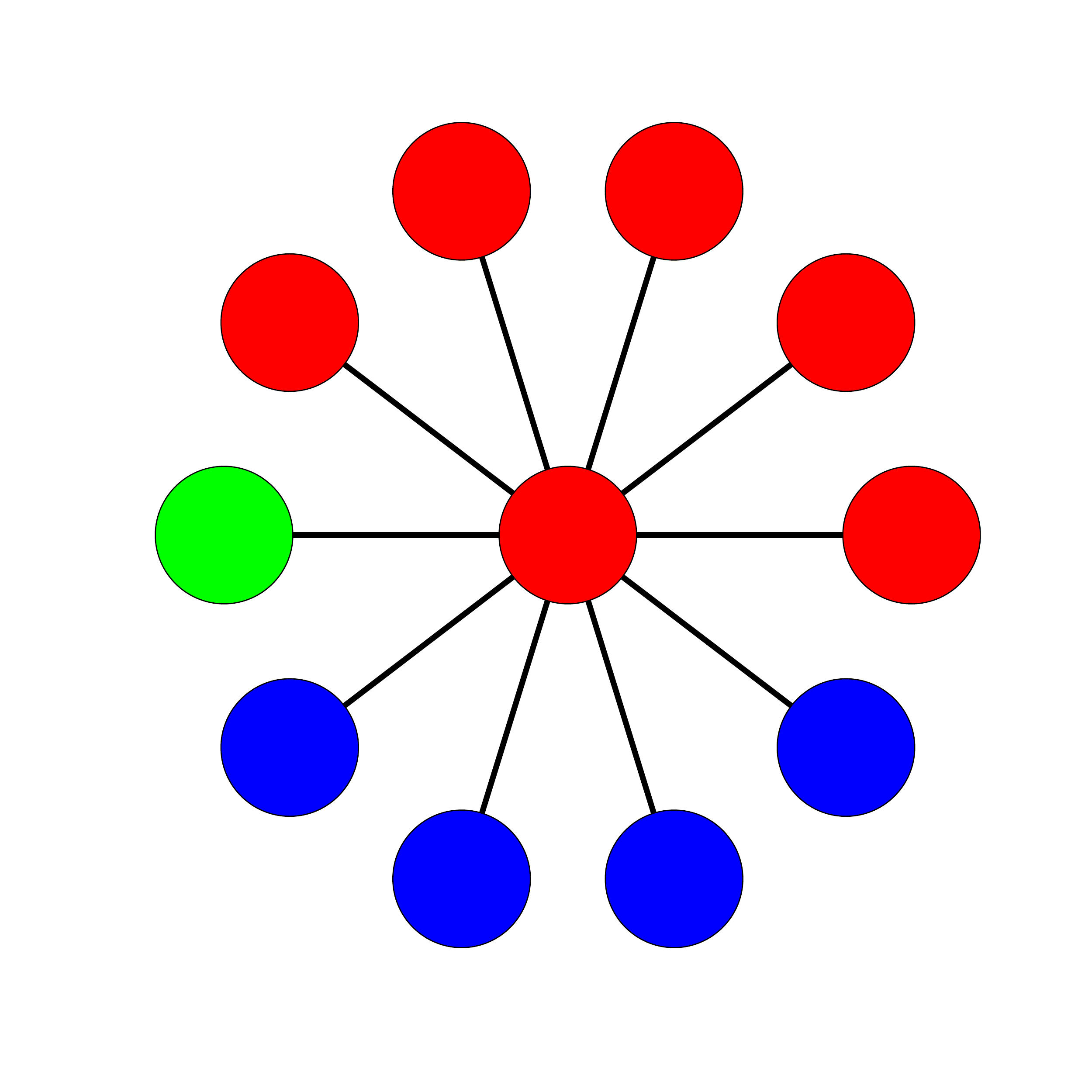}}
	\end{minipage}
	\begin{minipage}{0.32 \linewidth}
		\centerline{\includegraphics[scale=0.10]{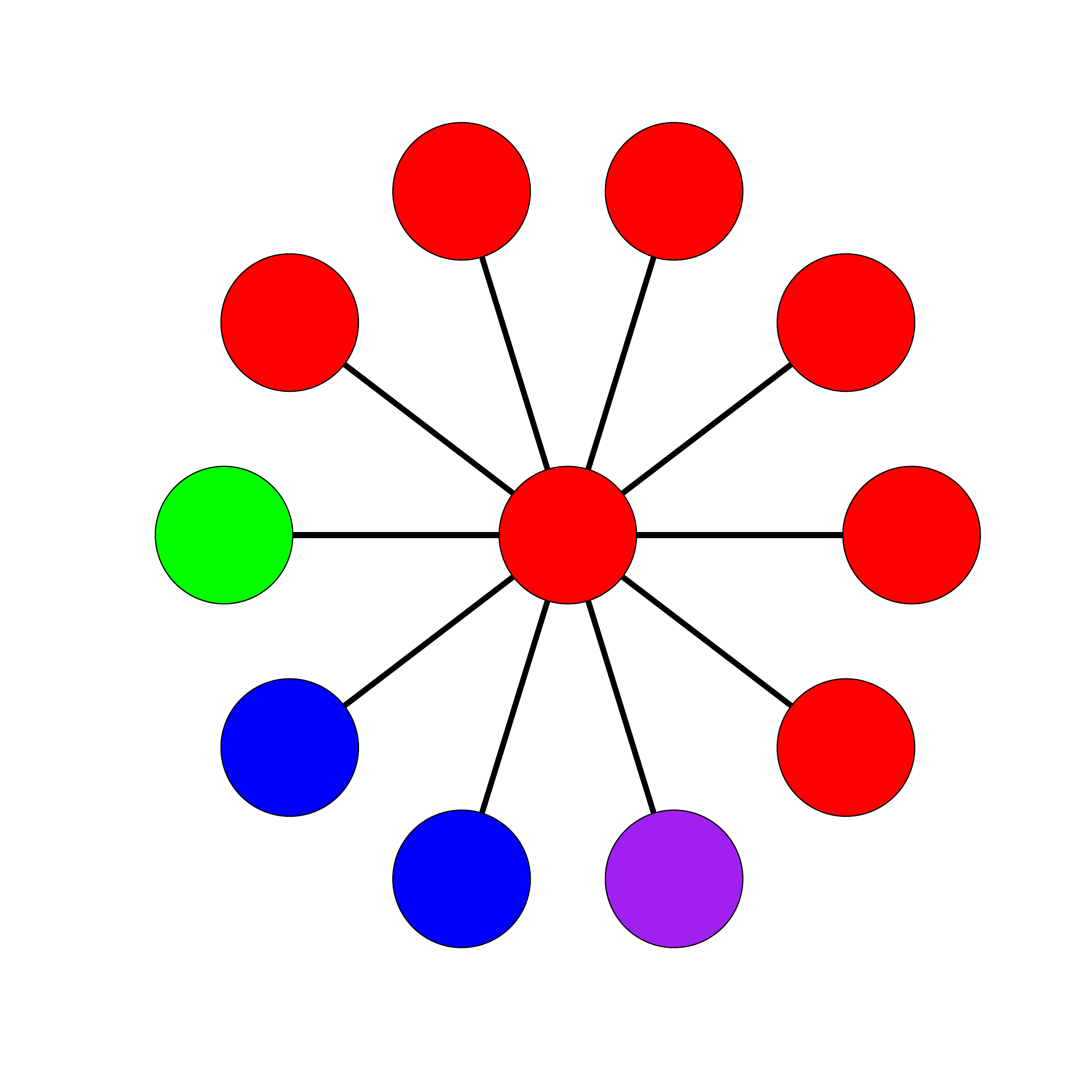}}
	\end{minipage}
	
		\caption{Each pattern represents a community. In each case, the participation of the central node is $0.58$.  \label{fig:participation}}  %TODO VL garder patterns ? ou revenir à couleurs ?
\end{figure}

\noindent \textbf{External Connectivity.}
\label{subsec:participationlimits}
We claim that the external connectivity of a given node, i.e. the way it is connected to communities other than its own, can be precisely described in three ways: first, by considering its \textit{diversity}, i.e. the number of concerned communities ; second, in terms of \textit{intensity}, i.e. the number of external links ; third relatively to its \textit{heterogeneity}, i.e. the distribution of external links over communities. The participation coefficient combines several of these aspects, mainly focusing on heterogeneity, which lowers its discriminant power. This is illustrated in Figure~\ref{fig:participation}: the external connectivity of the central node is very different in each one of the presented situations. However, $P$ is the same in all cases.

%As we shall see later, the number of external links of a node and the way its external links are distributed over communities are highly correlated in our data. More precisely, only nodes with a low number of external links are homogeneously connected to external communities, whereas nodes with many links are connected heterogeneously. Because of both this correlation and the definition of $P$, nodes with many external links will have a low participation coefficient in our experiments, which is not consistent with our aim. %TODO VL j'comprends plus la fin
%TODO AP perso j'ai jamais rien capte au paragraphe entier :) je vire

In order to be more illustrative, let us consider two users from our data, which have the same community role according to the original measures. We select two nodes both having a $z$ greater than $2.5$ and a $P$ close to $0.25$. So according to Guimer\`a and Amaral~\cite{Guimera2005} (see Table~\ref{tab:roleDesc}), they both are provincial hubs, and should have a similar behavior w.r.t. the community structure of the network. However, let us now point out that the first user is connected to $50$ nodes outside its community, whereas the second one has $200,000$ connections. This means they actually play different roles in the community structure, either because the second one is connected to much more communities than the first one, or because its number of links towards external communities is much larger than for the first user. Similar observations can be made for the directed variants of the participation coefficient. The measures used to define the external connectivity should take this difference into account and assign different roles to these nodes.%

\noindent \textbf{Fixed Thresholds.}
\label{subsec:threshold}
As indicated in the supplementary discussion of Guimer\`a and Amaral~\cite{Guimera2005}, the thresholds originally used to identify the roles were obtained empirically. They first processed $P$ and $z$ for different types of data: metabolic, proteome, transportation, collaboration, computer and random networks. Then, they detected basins of attraction, corresponding to regularities observed over all the studied networks. Each role mentioned earlier corresponds to one of these basins, and the thresholds were obtained by estimating their boundaries. 

Implicitly, these thresholds are supposed to be universal, but this can be criticized. First, Guimer\`a and Amaral~\cite{Guimera2005} used only one community detection method. A different community detection method can lead to a different community structure, and therefore possibly different basins of attraction. Furthermore, $z$ is not normalized, in the sense it has no fixed boundaries. There is no guarantee the threshold originally defined for this measure will stay meaningful on other networks. The values obtained for $z$ in our experiments are far higher for some nodes than the ones observed by~Guimer\`a and Amaral~\cite{Guimera2005}. We also observe that the proportion of nodes considered as hubs (i.e. $z \geq 2.5$) by~Guimer\`a and Amaral~\cite{Guimera2005} is much smaller in our network than in the networks they consider: $0.35\%$ in ours versus $2\%$ in theirs. These thresholds seem to be at least sensitive either to the size of the data, the structure of the network, or to the community detection method.

It is therefore necessary to process new thresholds, more appropriate to the considered data. However, the method used by Guimer\`a and Amaral~\cite{Guimera2005} itself is difficult to apply, because it requires a lot of data. We now present how to overcome these limitations. 

\subsection{Proposed Approach}
\label{sec:proposed}
%In this section, we propose some solutions to overcome the limitations of the original approach. First, the participation coefficient mixes several aspects of the external connectivity, which lowers its discriminant power: we introduce several measures to represent these aspects separately. Second, the thresholds used to define the roles do not necessarily hold for any system: we show how to apply an unsupervised method instead.

\noindent \textbf{Generalized Measures.}
\label{subsec:generalizedmeasures}
In place of the single participation coefficient, we propose $3$ new measures aiming at representing separately the aspects of external connectivity: \emph{diversity}, \emph{intensity} and \emph{heterogeneity}. %The first measure is related to the number of external communities connected to the node, the second to the number of its external links, and the last to the distribution of external links over communities. 
Moreover, a fourth measure is used to describe the internal connectivity. 

Because we deal with directed links, each one of these measures exists in two versions: incoming and outgoing (as explained in section \ref{sec:directedmeasures}), resulting in $8$ effective measures. However, for simplicity matters, we ignore link directions when presenting them in the rest of this section.

All our measures are expressed as $z$-scores. We know community sizes are generally power-law distributed, as described in~\cite{Lancichinetti2010}, which means their sizes are heterogeneous. Our community-based $z$-scores (cf. Equation (\ref{eq:zscore})) allows to normalize the measures relatively to the community size, and therefore to take this heterogeneity into account.

\noindent \emph{Diversity.} The \emph{diversity} $D(u)$ evaluates the number of communities to which a node $u$ is connected (other than its own), w.r.t. the other nodes of its community. This measure does not take into account the number of links $u$ has to each community. Let $\epsilon(u)$ be the number of external communities to which $u$ is connected. The diversity is defined as the $z$-score of $\epsilon$ w.r.t. the community of $u$. It is thus obtained by substituting $\epsilon$ to $f$ in Equation~(\ref{eq:zscore}).

\noindent \emph{External intensity.} The \emph{external intensity} $I_{ext}(u)$ of a node $u$ measures the amount of links $u$ has towards communities other than its own, w.r.t. the other nodes of its community. Let $d_{ext}(u)$ be the external degree of $u$, that is the number of links $u$ has with nodes belonging to another community than its own. The external intensity is defined as the $z$-score of the external degree, i.e. we obtain it by substituting $d_{ext}$ to $f$ in Equation~(\ref{eq:zscore}). 

\noindent \emph{Heterogeneity.} The \emph{heterogeneity} $H(u)$ of a node $u$ measures the variation of the number of links a node $u$ has, from one community to another. To that aim, we compute the standard deviation of the number of links $u$ has to each community. We denote this value by $\delta(u)$. The heterogeneity is thus the $z$-score of $\delta$ w.r.t. the community of $u$. As previously, it can be obtained by substituting $\delta$ to $f$ in Equation~(\ref{eq:zscore}).

\noindent \emph{Internal intensity.} In order to represent the internal connectivity of the node $u$, we use the $z$ measure of Guimer\`a and Amaral~\cite{Guimera2005}. Indeed, it is based on the notion of $z$-score, and is thus consistent with our other measures. Moreover, we do not need to add measures such as diversity or heterogeneity, since we consider one node can belong only to one community. Due to the symmetry of this measure with the external intensity, we refer to $z$ as the \emph{internal intensity}, and denote it by $I_{int}(u)$. 

\noindent \textbf{Unsupervised Role Identification.}
\label{sec:unsupervised}		
Our second modification concerns the way roles are defined. As mentioned before, the thresholds defined by Guimer\`a and Amaral~\cite{Guimera2005} are not necessarily valid for all data. Moreover, our generalization of the measures invalidates the existing thresholds, since we have now $8$ distinct measures, all different from the original ones. We could try estimating more appropriate thresholds, but as explained in section \ref{subsec:threshold}, the method originally used by Guimer\`a and Amaral~\cite{Guimera2005} to estimate their thresholds is impractical. The fact our measures are all $z$-scores also weakens the possibility to get thresholds applicable to all systems, which means the estimation process should potentially be performed again for each studied system.

To overcome these problems, we propose to apply an automatic method instead, by using unsupervised classification. First, we process all the measures for the considered data. Then, a cluster analysis method is applied. Each one of the clusters identified in the measure space is considered as a community role. This method is not affected by the number of measures used, and amounts to adjusting thresholds to the studied system. If the number of roles is known in advance, for instance because of some properties of the studied system, then one can use an appropriate clustering method such as $k$-means, which allows specifying the number $k$ of clusters to find. Otherwise, it is possible to use cluster quality measures to determine which $k$ is the most appropriate ; or to apply directly a method able to estimate at the same time the optimal number of clusters and the clusters themselves.

\section{Community Roles of Social Capitalists}
\label{sec:results}		
%In this section, we present the results we obtained on a Twitter network using the various methods presented in sections \ref{sec:original} and \ref{sec:proposed}. We first introduce the data and tools we used, then the roles we identified. 
%We then focus on the notion of social capitalist. We define this specific type of user present in Twitter, and describe how the identified roles can help understanding their position in the social network.

\subsection{Data and Tools}

We analyze a freely-available anonymized Twitter network, collected in 2009 by Cha \textit{et al}.~\cite{CHBG10}. It contains about $55$ million nodes representing Twitter users, and almost $2$ billion directed links corresponding to friend-to-follower relationships. We had to consider the size of these data when selecting our analysis tools. For community detection, we selected the Louvain method \cite{Blondel2008}, because it is widespread and proved to be very efficient when dealing with large networks. We retrieved the C++ source code published by its authors, and adapted it in order to optimize the directed version of the modularity measure, as defined by Leicht and Newman~\cite{Newman2008}. All the role measures, that is Guimer\`a and Amaral's original measures, their directed variants (section \ref{subsec:original}) and our new measures (section \ref{sec:proposed}), were computed using the community structure detected through this means. We also implemented them in C++, using the same sparse matrix data structure than the one used in the Louvain method. 

All resulting values were normalized, in order to avoid scale difference problems when conducting the cluster analysis. Since we do not know the expected number of roles, the clustering was performed using an open source implementation of a distributed version of $k$-means \cite{Liao2009}. Indeed, centralized versions are based on a unique distance matrix, and turned out to be too demanding in terms of memory. We applied this algorithm for $k$ ranging from $2$ to $15$, and selected the best partition in terms of Davies-Bouldin index~\cite{Davies1979}. We selected this index because it is a good compromise between the reliability of the estimated quality of the clusters, and the computing time it requires. All pre- and post-processing scripts related to the cluster analysis were implemented in R. 
The whole source code is available at the following address:  
\texttt{\url{https://github.com/CompNet/Orleans}} %TODO VL pour le package hyperref
%\texttt{https://github.com/CompNet/Orleans}.
%\texttt{https:...}.

\subsection{Roles Expected for Social Capitalists} 
\label{subsec:expectedrole}	
%We first discuss the roles we expect the social capitalists to hold. To estimate them, we use their degree and the previously mentioned ratio indices~\cite{DP14} to classify their behavior. 

We expect the degree of social capitalists to play an important role considering their position (see Section~\ref{sec:capsoc}). High in-degree social capitalists (namely greater than $10,000$) should be well connected to their communities -hubs- or to the other communities -connectors, or both. Being connectors would indicate they obtained a high visibility on the whole network and not only in their own communities.
Furthermore, because we take the direction of links into account in our measures, we expect social capitalists to be discriminated according to their ratio, i.e. the number of outgoing links divided by the number of incoming links. We especially expect high in-degree social capitalists with a small ratio (so-called \emph{passive social capitalists} according to~\cite{DP14}) to be highly connected to their communities and to the rest of the graph.
Considering low degree social capitalists, it is not possible to predict their roles without any further information. The study will thus be of great interest to characterize their visibility.

%The main issue lies in the notion of participation coefficient, which does not seem discriminative enough. Indeed, let us consider two nodes playing the same role in the community structure. More precisely, we consider hubs having a similar participation coefficient, that is users with a $z$-score greater than $2.5$ and a partition coefficient close to $0.25$. When using the repartition of roles proposed by~Guimer\`a and Amaral~\cite{Guimera2005}, those users are considered as provincial hubs. Hence, they should have a similar behavior w.r.t. the community structure of the network. However, we can observe that this is actually not the case: the first user is connected to $50$ users outside its community, while the second one is connected to $200000$. This means that they apparently play different roles in the community structure, either because the second one is connected to much more communities than the first one, or because its number of links towards external communities are much larger than the first user.

%We would like to notice that such measures are defined for undirected networks only, whereas the network we consider directed networks. This could be an explanation for this non accurate repartition. However, as we shall see Section~\ref{}, considering link direction within these measures will not improve the repartition. 

\subsection{Detected Roles}
For the sake of completeness, we first used the original undirected measures of Guimer\`a and Amaral~\cite{Guimera2005}. 
%As supposed in Section \ref{sec:proposed}, the threshold for $z$ turns out to be irrelevant, because this measure reaches much higher values than in the networks studied in Guimer\`a and Amaral~\cite{Guimera2005}. 
We obtained only $2$ roles, each one concerning too many nodes to bring up any valuable information regarding the studied system. Since this might be due to the fact these measures ignore link directions, we then worked with their directed variants (section \ref{subsec:original}), and then with our generalized measures (section \ref{sec:proposed}). In both cases we used the unsupervised role identification method we proposed (section~\ref{sec:unsupervised}).

\noindent \textbf{Directed Variants.}
%TODO CA SERAIT BIEN DE DIRE CE QU'ON TROUVE AVEC LEURS SEUILS AVANT DE PARTIR SUR LE CLUSTER ANALYSIS.
%TODO VL on n'a pas fait l'analyse avec les seuils, car ils sont aussi inconsistants par rapport aux valeurs qu'avec les mesures originales (on dit déjà dans la partie III que les seuils ne marchent ni pour les mesures originales ni pour les variantes orientées)
A correlation study shows $z^{out}$ and $z^{in}$ are slightly correlated (with a correlation coefficient $\rho<0.3$), whereas the correlation is zero for all other pairs of measures. This seems to confirm the interest of considering link directions in the role measures. When doing the cluster analysis, the most separated clusters are obtained for $k = 6$. 
%TODO AP on peut expliquer un peu ici peut-etre ? (je comprends aucun mot perso ^^)
An \textsc{ANOVA} followed by \textit{post hoc} tests ($t$-test with Bonferroni's correction) showed significant differences exist between all clusters and for all measures. 
%
%\begin{table}[h]
%	\centering
%	\begin{tabular}{|l|r|r|r|}
%		\hline
%		\textbf{C} & \textbf{Size} & \textbf{\%} & \textbf{Role} \\
%		\hline
%		1 & $3352534$ & $6.38$ & In. Kinless non-hubs \\
%		  &  &  & Out. U.-periph. non-hubs \\
%		\hline
%		2 &      $420$ & $<0.01$ & Connector hubs \\
%		\hline
%		3 &   $15510589$ &  $29.50$ & Periph. non-hubs	\\
%		\hline
%		4 & $7868064$ & $14.96$ & In. U.-periph. non-hubs \\
%		  &  &  & Out. Kinless non-hubs \\
%		\hline
%		5 & $13936680$ & $26.51$ & U.-periph. non-hubs \\
%		\hline
%		6 &  $11911395$ &  $22.65$ & In. Periph. non-hubs \\
%		  &   &   & Out. U.-periph. non-hubs \\
%		\hline
%	\end{tabular}
%	\caption{Clusters detected with the directed measures: sizes in terms of node count and proportion of the whole network, and roles according to the Guimer\`a and Amaral~\cite{Guimera2005} nomenclature.}
%	\label{tab:groupes_directed}
%\end{table}

An analysis of the distribution of high in-degree social capitalists in these clusters shows that a few of these users occupy a connector hub role. This is quite expected as said in~\ref{subsec:expectedrole}. However, most of the high degree social capitalists are considered as non-hubs and peripheral or ultra-peripheral nodes. More than $60\%$ of the users with a high ratio are classified as ultra-peripheral nodes for both incoming and outgoing directions, which is rather surprising since they have a really high degree. However, they are classified in a cluster with low $z$ and $P$ (both in- and out- versions). The low $z$ indicates these users are not much connected to their community (relatively to the other nodes of the same community), and must thus be more connected to other communities. Still, $P$ does not highlight this aspect of their community-related connectivity, and they appear as peripheral. This inconsistency of the detected roles confirms the limitations of $P$ described in section \ref{subsec:participationlimits}. %We will see that these problems do not appear with our generalized measures.

\noindent \textbf{Generalized Measures.}
Most generalized measures are slightly correlated, with values ranging from almost $0$ to $0.4$. In particular, both versions of the same measure (incoming vs. outgoing) are only slightly correlated, which is another confirmation of the interest of considering link directions. Only three measures are strongly correlated: internal and external intensities and heterogeneity ($\rho$ ranging from $0.78$ to $0.92$). The relation between both intensities seems to indicate that variations on the total degree globally affect similarly internal and external degrees. The very strong correlation observed between heterogeneity and intensity means only nodes with low intensity are homogeneously connected to external communities, whereas nodes with many links are connected heterogeneously.

Similarly to the directed measures, the most separated clusters are obtained with $k = 6$. These $6$ clusters are given in Table~\ref{tab:groupes_generalized} with their sizes and roles. However, the correspondance with the original nomenclature is rougher, since these measures are farther from the original ones. The average of each measure per cluster is showed in Table~\ref{tab:moyennes_generalized}. Like before, \textsc{ANOVA} and \textit{post hoc} tests showed significant differences between all clusters and for all measures. We now conduct a detailed analysis of the different roles we obtain. 

\begin{table}[h]
	\centering
	\begin{tabular}{|c|r|r|r|}
		\hline
		\textbf{Cluster} & \textbf{Size} & \textbf{Proportion} & \textbf{Role} \\
		\hline
		1 & $24543667$ & $46.68\%$ & Ultra-peripheral non-hubs \\
		2 &      $304$ & $<0.01\%$ & Kinless hubs \\
		3 &   $303674$ &  $0.58\%$ & Connector hubs	\\
		4 & $11929722$ & $22.69\%$ & Incoming Peripheral non-hubs \\
		5 & $10828599$ & $20.59\%$ & Outgoing Peripheral non-hubs \\
		6 &  $4973717$ &  $9.46\%$ & Connector non-hubs \\
		\hline
	\end{tabular}\\[0.4cm]
	\caption{Clusters detected with the generalized measures: sizes in terms of node count and proportion of the whole network, and roles according to the Guimer\`a and Amaral~\cite{Guimera2005} nomenclature.}
	\label{tab:groupes_generalized}
\end{table}

\noindent \textit{Cluster 1.} Because both internal intensity versions (equivalent to $z$) are negative, nodes in this cluster cannot be hubs. The negative external measures indicate these nodes are not connectors either. We can thus consider them as ultra-peripheral non-hubs. This cluster is the largest one, with $47\%$ of the network nodes. This confirms the matching with this role, whose nodes constitute generally most of the network.

\noindent \textit{Clusters 4 and 5.} Cluster $4$ is very similar to Cluster $1$. However, its incoming diversity is $0.69$. These nodes are again peripheral, because the external intensity is negative. Still, incoming links come from a larger number of communities.
Cluster $5$ is also similar to Cluster $1$. However, both versions of diversity are positive for this cluster, with an outgoing diversity of $0.60$. External links are thus connected to a larger number of communities.
Clusters $4$ and $5$ are the second ($23\%$) and third ($21\%$) largest ones, respectively. By gathering all the peripheral and ultra-peripheral nodes, we obtain $91\%$ nodes of the network.

\noindent \textit{Cluster 6.} The internal intensity is still close to $0$ but positive. Thus, these nodes are non-hubs, even if they are more connected to their community than those of the previous clusters. Like the other external measures, the external intensity is low but still positive. These nodes are relatively well-connected to other communities, and we can therefore consider them as connectors. Both versions of the diversity are relatively high, which indicates these nodes are not only more connected to their community as well as others, but also to a larger number of distinct communities.

\noindent \textit{Cluster 3.} The high internal intensity allows us to state that these nodes are hubs. Furthermore, the high external measures indicate these nodes are connected to a high number of nodes from a lot of other communities, and thus are connector hubs. Notice outgoing measures are higher. This cluster represents only $0.6\%$ of the network, meaning this role is very uncommon. 

\noindent \textit{Cluster 2.} This observation is even more valid for Cluster $2$, which represents much less than $1\%$ of the nodes. For this cluster, all measures are really high. The incoming versions are always higher than their outgoing counterparts. We call these users kinless hubs according to Guimer\`a and Amaral's nomenclature.

\begin{table}[h]
	\centering
	\begin{tabular}{|c|r|r|r|r|}
		\hline
		$\mathbf{Cluster}$ & $\mathbf{I^{out}_{int}}$ & $\mathbf{I^{in}_{int}}$ & $\mathbf{D^{out}}$ & $\mathbf{D^{in}}$\\
		\hline
		1 & $-0.12$ &  $-0.03$ & $-0.55$ & $-0.80$ 	\\
		2 & $94.22$ & $311.27$ &  $7.18$ & $88.40$ 	\\
		3 &  $5.52$ &   $1.40$ &  $5.60$ &  $3.10$ 	\\
		4 & $-0.04$ &   $0.00$ & $-0.37$ &  $0.69$ 	\\
		5 & $-0.03$ &  $-0.01$ &  $0.60$ &  $0.19$ 	\\
		6 &  $0.48$ &   $0.12$ &  $1.96$ &  $1.70$ 	\\
		\hline
	\end{tabular}\\[0.2cm]
	\begin{tabular}{|c|r|r|r|r|}
		\hline
		$\mathbf{Cluster}$ &
 		$\mathbf{I^{out}_{ext}}$ & $\mathbf{I^{in}_{ext}}$ & $\mathbf{H^{out}}$ & $\mathbf{H^{in}}$ \\
		\hline
		1 & $-0.09$ &  $-0.04$ &  $-0.12$ &  $-0.06$ 	\\
		2 & $113.87$ & $283.79$ & $112.79$ & $285.57$ 	\\
		3 &  $5.28$ &   $1.43$ &   $6.76$ &   $2.34$ 	\\
		4 & $-0.07$ &   $0.00$ &  $-0.10$ &  $-0.01$ 	\\
		5 & $-0.03$ &  $-0.02$ &  $-0.04$ &  $-0.02$ 	\\
		6 & $0.35$ &   $0.12$ &   $0.53$ &   $0.19$ 	\\
		\hline
	\end{tabular}\\[0.4cm]
	
	\caption{Average generalized measures obtained for the $6$ detected clusters 
	\label{tab:moyennes_generalized}.}
\end{table}

It is worth noticing that, whatever the considered measures, some of the roles defined by Guimer\`a and Amaral~\cite{Guimera2005} are not represented in the studied network. This is consistent with the remarks previously made for other data by Guimer\`a and Amaral~\cite{Guimera2005}, and  confirms the necessity of having an unsupervised approach to define roles in function of measures. It is also consistent with the strong correlation observed between internal and external intensities: missing roles would be nodes possessing a high internal intensity but a low external one, or vice-versa. However, those are very infrequent in our network.
%Link directions constitute another interesting aspect. Some groups estimated with the directed or generalized measures are considered as equivalent by the original undirected measures. This means distinguishing incoming and outgoing links led to a finer typology. 
%TODO PARTICIPATION NE DONNE QU'UNE VISION FLOUE DES CONNEXIONS EXTERNES ?

\subsection{Relations between clusters}
\label{subsec:relations}
\begin{figure}
	\center	
	\includegraphics[scale=0.58]{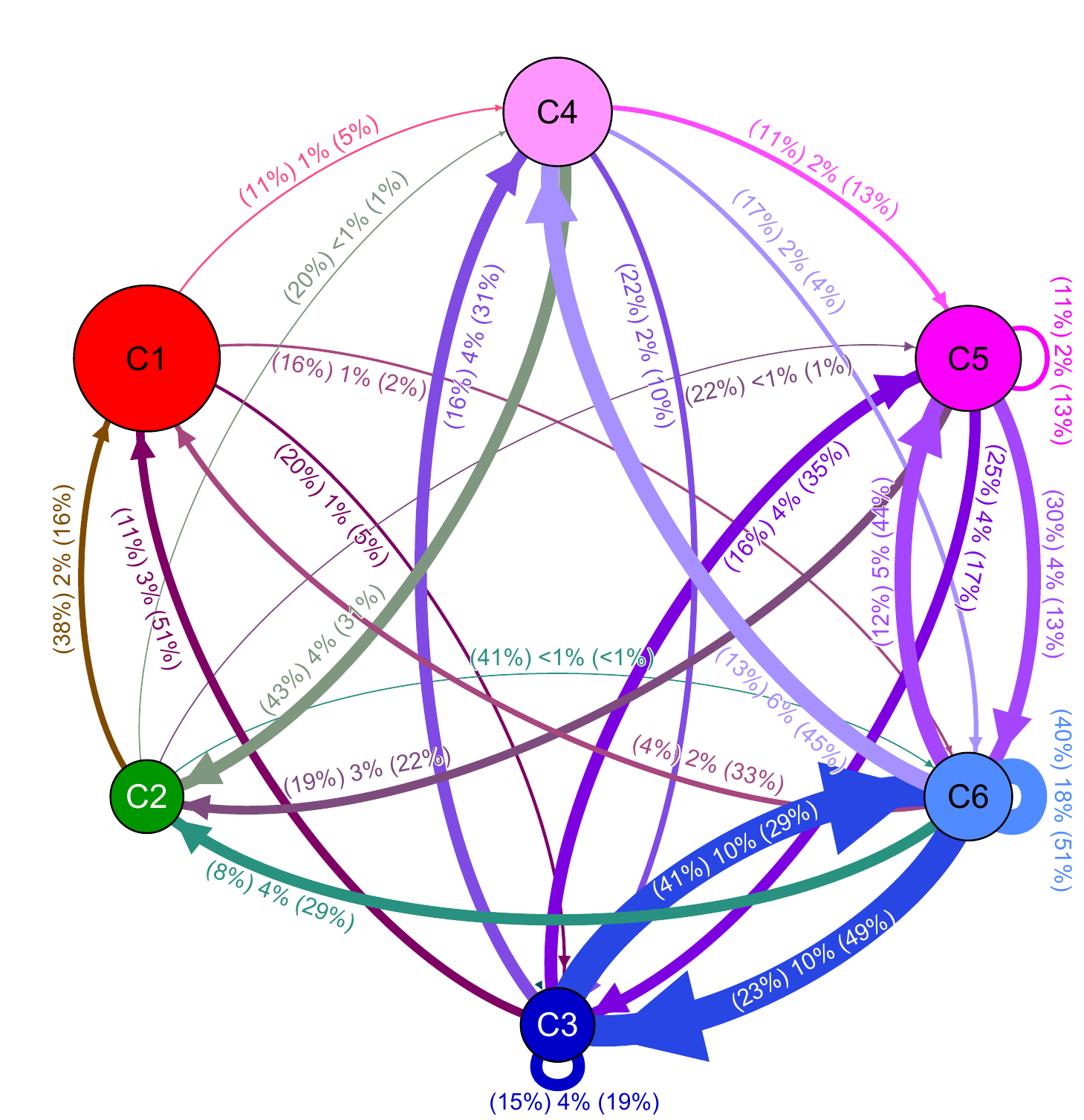}
	\caption{Interconnection between clusters. A vertex $i$ corresponds to Cluster $i$ from Table \ref{tab:groupes_generalized}. An arc $(i,j)$ represents the set of links connecting nodes from Cluster $i$ to nodes from Cluster $j$. It is labeled with 3 values, each one describing which proportion of links the arc represents, relatively to 3 distinct sets: first relatively to all links starting from Cluster $i$, second relatively to all links in the whole network, and third relatively to all links ending in Cluster $j$. The arc thickness is proportional to the second value. For matters of readability, arcs representing less than $1\%$ of the network links are not displayed.}\label{fig:ctoc}
\end{figure}

We now discuss how the nodes are connected depending on the role they hold. Figure~\ref{fig:ctoc} is a simplified representation of this interconnection pattern. 
%Each vertex corresponds to one of the $6$ clusters identified in the previous section, and each arc represents a set of links between nodes holding two particular roles. Note the arcs representing less than $1\%$ of the network links and $10\%$ of the cluster links are not displayed.

The outgoing links of ultra-peripheral (Cluster 1) and peripheral (Clusters 4 and 5) nodes target mainly kinless hubs (Cluster 2) and connectors (Clusters 3 and 6), representing $74\%$ (Cluster 1), $82\%$ (Cluster 4), and $74\%$ (Cluster 5) of their connections. These (ultra-)peripheral nodes, which are the most frequent in the network, thus mainly follow very connected users, probably the most influent and relevant ones. This seems consistant: they follow only a few users, and so choose the most visible ones. 
%A similar remark can be made for the incoming links, but this time only for connectors (Cluster 3 and Cluster 6), which represent $84\%$ (Cluster 1), $76\%$ (Cluster 4) and $79\%$ (Cluster 5) of their connections.

Connector nodes (Clusters 3 and 6) are mainly linked to other connectors nodes. They have the tightest connection, since their arcs amounts to a total of $43\%$ of the network links. This is worth noticing, because these clusters are far from being the largest ones. They are also largely connected to the rest of the clusters too, especially with outgoing links. 
%However, there is a difference between connector hubs and non-hubs, on this point: the latter are significantly connected to kinless hubs ($8\%$ of Cluster 6 outgoing links), which is not true at all for the former (only $1\%$).
Connectors follow massively users of all clusters, so we suppose they constitute the backbone of the network.

Kinless hubs (Cluster 2) are massively followed by non-hubs, representing $38\%$ (Cluster 1), $43\%$ (Cluster 4), $19\%$ (Cluster 5) and $8\%$ (Cluster 6) of their outgoing links. And interestingly, the links coming from kinless hubs target the same clusters: $9\%$ go to Cluster 1, $20\%$ to Cluster 4, $22\%$ to Cluster 5 and $41\%$ to Cluster 6. This means the most visible and popular nodes of the network mostly follow and are followed by much less popular users. One could have expected the network to be hierarchically organized around roles, with more peripheral nodes connected to less peripheral nodes. But this is clearly not the case. First, (ultra-)peripheral nodes are marginally connected to other nodes holding the same role, they prefer to follow connectors and/or hubs. Second, kinless and connector hubs, although well connected to connector non-hubs, do not have direct links, i.e. these users do not follow each other.

%\begin{figure}
%	\includegraphics[scale=0.65]{figures/ctoc_out}
%	\caption{Each node represent the nodes of one cluster, the node id is the number of this cluster. The weight of an arc $(i,j)$ represents the percentage of outgoing links of $i$ directed to $j$. To be clearer, only the most meaningful arcs are considered, namely those with the higher weights.}\label{fig:ctoc_out}
%\end{figure}

%\begin{figure}[h]
%	\includegraphics[scale=0.55]{figures/ctoc_in}
%	\caption{Each node represent the nodes of one cluster, the node id is the number of this cluster. The weight of an arc $(i,j)$ represents the percentage of incoming links of $i$ coming from $j$. To be clearer, only the most meaningful arcs are considered, namely those with the higher weights.}\label{fig:ctoc_in}
%\end{figure}

\subsection{Position of Social Capitalists}
\label{sec:position}
As stated previously, we use a list of approximately $160,000$ social capitalists as detected by Dugu\'e and Perez~\cite{DP14}. In the following, we analyze how social capitalists are distributed amongst the detected roles. As explained section~\ref{sec:capsoc}, we split social capitalists according to their in-degree (number of followers). Recall that \emph{low in-degree social capitalists} have an in-degree between $500$ and $10,000$, and \emph{high in-degree social capitalists} an in-degree greater than $10,000$. These social capitalists are known for having especially well succeeded in their goal of gaining visibility.

The tables in this section describe how the various types of social capitalists are distributed over the clusters.
In each cell, the first row is the proportion of social capitalists belonging to the corresponding cluster, and the second one is the proportion of cluster nodes which are social capitalists. Values of interest are indicated in bold and discussed in the text.

\noindent \textbf{Low in-degree social capitalists.}
Low in-degree social capitalists are mostly assigned to three clusters: $3$, $5$ and $6$ (see Table~\ref{tab:ksociaux500_generalized}). Most of them belong to Cluster $6$, which contains non-hub connector nodes. These nodes, which have only slightly more external connections than the others, are nevertheless connected to far more communities. Social capitalists in this cluster seem to have applied a specific strategy consisting in creating links with many communities. This strategy is still not completely working, though, as shown by the relatively low external incoming intensity (meaning they do not have that many followers). %TODO VL j'ai tenté de clarifier cette phrase, j'espère ne pas avoir introduit d'erreur. effacez ce commentaire si c'est ok.

Nodes from Cluster $3$ are connector hubs, who follow more users than the others. Because \textbf{IFYFM} social capitalists have a ratio greater than $1$ and thus more friends than followers, it is quite intuitive to observe that they are twice as many than the other users in this cluster. The high outgoing diversity of Cluster $3$ tells us that these social capitalists follow users from a large variety of communities, not only theirs (to which they are well connected). The high external outgoing intensity show that these users massively engage in the \textbf{IFYFM} process, but did not yet receive a lot of following back, as shown by their low external incoming intensity.
Finally, roughly $20\%$ of social capitalists with ratio $r < 1$ belong to Cluster $5$, which contains non-hub peripheral nodes. This shows that a non-neglictible share of social capitalists are isolated relatively to both their community and the other ones. 

\begin{table}[h]
	\centering
	\begin{tabular}{|l|r|r|r|r|r|r|}
		\hline
		\textbf{Ratio} & \textbf{Cluster 1}  & \textbf{Cluster 2} & \textbf{Cluster 3} \\
		\hline
	  	\multirow{2}{*}{$r \leq 1$} & $0.01\%$ & $0.00\%$ & $\mathbf{23.10\%}$   \\
		& $< 0.01\%$ & $0.00\%$ & $3.71\%$    \\
		\hline
		\multirow{2}{*}{$r > 1$}  & $0.03\%$   & $0.00\%$ & $\mathbf{18.78\%}$   \\
		& $< 0.01\%$ & $0.00\%$ & $\mathbf{6.61}\%$  \\
		\hline
	\end{tabular}\\[0.2cm]
	\begin{tabular}{|l|r|r|r|r|r|r|}
		\hline
		\textbf{Ratio}  & \textbf{Cluster 4} & \textbf{Cluster 5} & \textbf{Cluster 6}\\
		\hline
		\multirow{2}{*}{$r \leq 1$}   &  $3.42\%$ &  $\mathbf{18.28\%}$ & $\mathbf{55.19\%}$  \\
		&  $0.14\%$ &  $0.08\%$ & $0.54\%$  \\
		\hline
		\multirow{2}{*}{$r > 1$}    &  $0.48\%$   &  $\mathbf{14.31\%}$ & $\mathbf{66.40\%}$ \\
		& $< 0.01\%$ &   $0.14\%$ 				& $1.43\%$ \\
		\hline
	\end{tabular}\\[0.4cm]
	\caption{Distribution of low in-degree social capitalists over clusters obtained from the generalized measures.
	\label{tab:ksociaux500_generalized}}
\end{table}

These observations show that most of these users are deeply engaged in a process of soliciting users from other communities, not only theirs. Some of them are even massively following users from a wide diversity of communities. This tends to show that these users may obtain an actual visibility across many communities of the network by spreading their links efficiently.

\noindent \textbf{High in-degree social capitalists.}
Most of the high in-degree social capitalists are gathered in Cluster $3$ (see Table~\ref{tab:ksociaux10000_generalized}), corresponding to connector hubs. This is consistent with the fact these users have a high degree. Users of Cluster $3$ have a high outgoing diversity and a high outgoing external intensity: this shows they practice the \textbf{IFYFM} strategy actively, by following a lot of users from a wide range of communities.
The rest of these users is contained in Cluster $2$. Nodes in these clusters are kinless hubs and thus can be considered as successful users. Indeed, they are massively followed by a very high number of users from an extremely large variety of communities.
Only high-degree social capitalists with a ratio smaller than $0.7$ and a few with a ratio smaller than $1$ are classified in this cluster. This is consistent with the roles one could expect for social capitalists (section~\ref{sec:capsoc}).

\begin{table}[h]
	\centering
	\begin{tabular}{|l|r|r|r|}
		\hline
		\textbf{Ratio} & \textbf{Cluster 1}  & \textbf{Cluster 2} & \textbf{Cluster 3} \\
		\hline
		\multirow{2}{*}{$r \leq 0.7$} & $0.00\%$ & $\mathbf{12.14\%}$ & $\mathbf{87.29\%}$ \\
		& $0.00\%$ & $\mathbf{21.05\%}$ & $0.15\%$ \\
		\hline
		\multirow{2}{*}{$0.7 < r \leq 1$} & $0.00\%$ & $1.55\%$ & $\mathbf{95.64\%}$ \\
		& $0.00\%$ & $\mathbf{7.24\%}$ & $0.45\%$ \\
		\hline
		\multirow{2}{*}{$r > 1$} & $0.00\%$ & $0.03\%$ & $\mathbf{97.99\%}$ \\
		& $0.00\%$ & $0.33\%$ & $1.22\%$ \\
		\hline
	\end{tabular}\\[0.2cm]
	\begin{tabular}{|l|r|r|r|}
		\hline
		\textbf{Ratio} & \textbf{Cluster 4} & \textbf{Cluster 5} & \textbf{Cluster 6}\\
		\hline
		\multirow{2}{*}{$r \leq 0.7$}  & $0.00\%$  & $0.00\%$ & $0.57\%$ \\
		&  $0.00\%$ & $0.00\%$ & $< 0.01\%$ \\
		\hline
		\multirow{2}{*}{$0.7 < r \leq 1$}  & $0.00\%$ & $0.00\%$ & $2.81\%$ \\
		& $0.00\%$ &  $0.00\%$ & $< 0.01\%$ \\
		\hline
		\multirow{2}{*}{$r > 1$}  & $0.00\%$ & $0.00\%$ & $1.98$ \\
		& $0.00\%$ & $0.00\%$ & $< 0.01\%$ \\
		\hline
	\end{tabular}\\[0.4cm]
	\caption{Distribution of high in-degree social capitalists over clusters obtained from the generalized measures.
	\label{tab:ksociaux10000_generalized}}
\end{table}
These observations mean that most of these users are well connected in their communities but also with the rest of the network. This shows the efficiency of these users strategies. Indeed, most of the users are linked to a wide range of communities, and thus reach a high visibility in a large part of the network.

\section{Conclusion}
\label{sec:conclusion}		
In this article, our goal is to characterize the position of social capitalists in Twitter. For this purpose, we propose an extension of the method defined by Guimer\`a and Amaral~\cite{Guimera2005} to characterize the community role of nodes in complex networks. We first define directed variants of the original measures, and extend them further in order to take into account the different aspects of node connectivity. Then, we propose an unsupervised method to determine roles based on these measures. It has the advantage of being independant from the studied system. Finally, we apply our tools to a friend-to-follower Twitter network. We find out the different kinds of social capitalists occupy very specific roles. Those of low in-degree are mostly connectors non-hubs. This shows they are engaged in a process of spreading links across the whole network, and not only their own community. Those of high in-degree are classified as kinless or connectors hubs, depending on their ratio $r$. This shows the efficiency of their strategies, which lead to a high visibility for a vast part of the network, not only for their own community.

The most direct perspective for our work is to assess its robustness. In particular, it is important to know how the stability of the detected communities and clusters affects the identified roles. In this study, the very large size of the data prevented us to do so: first, it was a strong constraint when selecting the tools we used for community detection and cluster analysis, and second it was not possible to repeat these processing many times to evaluate the stability of their results. We plan to work on this point by using smaller datasets. On a related note, we want to apply our method to other systems, in order to check for its general relevance. The method itself can also be extended in two ways. First, it would be relatively straightforward to take link weights into account (although this was not needed for this work). Second, and more interestingly, it is also possible to adapt it to overlapping communities (by opposition to the mutually exclusive communities considered in this work) in a very natural way, by introducing additional internal measures symmetrical to the existing external ones. This could be a very useful modification when studying social networks, since those are supposed to possess this kind of community structures, in which a node can belong to several communities at once \cite{Arora2012}.

\bibliographystyle{IEEEtran}
\bibliography{biblio_ap-lg}

\end{document}